\documentclass[conference]{IEEEtran}

\usepackage{cite}
\usepackage{amsmath,amssymb,amsfonts}
\usepackage{algorithmic}
\usepackage{graphicx}
\usepackage{textcomp}
\usepackage{xcolor}
\usepackage{tikz}
\usepackage{hyperref}
\usetikzlibrary{positioning, arrows, shapes.geometric}

\def\BibTeX{{\rm B\kern-.05em{\sc i\kern-.025em b}\kern-.08em
    T\kern-.1667em\lower.7ex\hbox{E}\kern-.125emX}}

\makeatletter
\newcommand{\linebreakand}{%
    \end{@IEEEauthorhalign}
    \hfill\mbox{}\par
    \mbox{}\hfill\begin{@IEEEauthorhalign}
}
\makeatother

\begin{document}

\title{Crowd-SFT: Crowdsourcing for LLM Alignment}

\author{\IEEEauthorblockN{Alex Sotiropoulos}
\IEEEauthorblockA{\textit{Viterbi School of Engineering} \\
\textit{University of Southern California}\\
Los Angeles, USA \\
asotirop@usc.edu}
\and
\IEEEauthorblockN{Sulyab Thottungal Valapu}
\IEEEauthorblockA{\textit{Viterbi School of Engineering} \\
\textit{University of Southern California}\\
Los Angeles, USA \\
thottung@usc.edu}
\and
\IEEEauthorblockN{Linus Lei}
\IEEEauthorblockA{\textit{Viterbi School of Engineering} \\
\textit{University of Southern California}\\
Los Angeles, USA \\
linuslei@usc.edu}
\linebreakand
\IEEEauthorblockN{Jared Coleman}
\IEEEauthorblockA{\textit{Seaver College of Science and Engineering} \\
\textit{Loyola Marymount University}\\
Los Angeles, USA \\
jared.coleman@lmu.edu}
\and
\IEEEauthorblockN{Bhaskar Krishnamachari}
\IEEEauthorblockA{\textit{Viterbi School of Engineering} \\
\textit{University of Southern California}\\
Los Angeles, USA \\
bkrishna@usc.edu}
}

\maketitle

\IEEEpeerreviewmaketitle

\begin{abstract}
Large Language Models (LLMs) increasingly rely on Supervised Fine-Tuning (SFT) and Reinforcement Learning from Human Feedback (RLHF) to align model responses with human preferences. While RLHF employs a reinforcement learning approach with a separate reward model, SFT uses human-curated datasets for supervised learning. Both approaches traditionally depend on small, vetted groups of annotators, making them costly, prone to bias, and limited in scalability. We propose an open, crowd-sourced fine-tuning framework that addresses these limitations by enabling broader feedback collection for SFT without extensive annotator training. Our framework promotes incentive fairness via a point-based reward system correlated with Shapley values and guides model convergence through iterative model updates. Our multi-model selection framework demonstrates up to a 55\% reduction in target distance over single-model selection, enabling subsequent experiments that validate our point-based reward mechanism's close alignment with Shapley values—a well-established method for attributing individual contributions—thereby supporting fair and scalable participation.
\end{abstract}

\begin{IEEEkeywords}
SFT, Large Language Models, Crowdsourcing, Tournament-Based Selection
\end{IEEEkeywords}

\section{Introduction}
As Large Language Models (LLMs) become integral to various applications, aligning their responses with human preferences is crucial.
Supervised Fine-Tuning (SFT) and Reinforcement Learning from Human Feedback (RLHF) have emerged as the standard approaches for accomplishing this.
Nonetheless, both SFT and RLHF face fundamental limitations: SFT requires curated, high-quality datasets that are typically expensive to create~\cite{wang2023aligninglargelanguagemodels}, and both approaches suffer from limited annotator diversity, introducing potential biases and scalability challenges~\cite{casper2023openproblemsfundamentallimitations, bai_training_2022}.
In this paper, we propose an \textit{online} SFT framework in which models are iteratively updated with new human feedback.
While our focus is on SFT, the same principles could be applied to RLHF by adapting them for reward model training.

We propose a framework that leverages a decentralized, crowd-sourced approach to fine-tuning. Instead of relying on a small set of evaluators, our method allows a broader pool of users to iteratively refine an LLM while tracking individual contributions. At each iteration, users are divided into non-overlapping groups, with each group collectively fine-tuning a separate model instance. The best-performing model, selected through an evaluation function, is used as the base for the next iteration. Although only one model advances, users in all groups receive rewards based on their group's performance relative to the baseline model, ensuring even non-winning groups are proportionately compensated. Our approach not only democratizes fine-tuning but also introduces a competitive dynamic that enhances model quality while maintaining fairness in rewards.

Our key contributions~\footnote{Code available at: \url{https://github.com/ANRGUSC/ml-bc-rating-review}} are:
\begin{itemize}
    \item We propose a novel iterative fine-tuning framework in which multiple user groups train competing model instances.
    A selection mechanism ensures continual model improvement across iterations, while a point-based system allocates rewards to participants proportionally to their group's contribution in each iteration.
    \item We empirically validate our framework through simulations, demonstrating improved model convergence and accurate estimation of group-wise contributions.
    \begin{itemize}
        \item In simulations, our competitive fine-tuning framework improved model convergence by as much as 55\% compared to traditional single-model SFT.
        \item The proposed point-based reward system successfully estimated individual user contributions, correlating with their actual impact as determined by Shapley values—a robust method from cooperative game theory for fairly attributing contributions to collective outcomes.
        \item We also observed that incorporating randomness in user grouping and model evaluation improved both convergence and fairness in reward distribution.
    \end{itemize}
\end{itemize}

The remainder of this paper is organized as follows: Section~\ref{sec:related_work} reviews related work, Section~\ref{sec:point_space} introduces a method to empirically measure LLM fine-tuning as movement in vector space, and lays the foundation for our collaborative fine-tuning framework. Expanding on this, Section~\ref{sec:grouping_eval_contr} details the framework and analyzes its performance and scaling effects across various grouping and evaluation methods. Section~\ref{sec:conclusion} concludes the paper, and finally, Section~\ref{sec:future} outlines future research directions.

\section{Related Work} \label{sec:related_work}
SFT calibrates pre-trained models by directly training them on high-quality instruction-response pairs, aligning LLMs with human preferences. While SFT focuses on demonstrative learning, it is often complemented by RLHF, which optimizes model behavior using comparative human feedback in a reinforcement learning framework~\cite{ouyang2022traininglanguagemodelsfollow}. Despite their effectiveness, approaches relying on human feedback face fundamental limitations in data collection due to small annotator pools, which significantly restrict alignment methods. Narrow evaluator demographics can embed political and cultural biases, while inconsistencies in human judgment—exacerbated by cognitive fatigue over extended annotation sessions—further weaken reliability~\cite{casper2023openproblemsfundamentallimitations, bai_training_2022}. These challenges suggest the need for a more open and inclusive SFT process that leverages broader human feedback. 

As the foundation of our approach, we utilize tournament-based selection, where user groups fine-tune separate model copies. The best-performing version becomes the baseline for the next round, driving improvement through competition. Tournament-based selection has been widely used to benchmark LLM performance through model comparisons~\cite{chiang2024chatbotarenaopenplatform, son2024varcoarenatournamentapproach}. However, our framework extends this paradigm by incorporating contribution-aware tournaments that evaluate model quality through pairwise battles and track user impact via a structured point-based system validated against Shapley values. 

Though not directly integrated into our approach, Shapley values serve as a baseline for evaluating our point-based system by fairly attributing individual contributions to collective outcomes. Initially developed in cooperative game theory, they have been widely applied in machine learning to quantify the importance of features, data points, and individual participants in collaborative settings. In the context of LLM training, Shapley values have been used to allocate rewards in online data marketplaces~\cite{agarwal_marketplace_2019} and to evaluate contributions in federated learning systems~\cite{ma_transparent_2021}. These prior works primarily focus on objective data sources, such as medical or sensor data; in contrast, our framework extends their application to subjective human feedback in the SFT setting.

\section{Experiment 1: Vector-Space Modeling of LLM Fine-Tuning} \label{sec:point_space}
We conceptualize LLM fine-tuning as an iterative process in which model updates are guided by user feedback. Our approach assumes that there exists some embedding function $\Phi$ that maps model outputs to an $n$-dimensional vector space, allowing us to quantify their alignment with a predefined target. While the actual fine-tuning process remains opaque, this assumption enables us to analyze the effects of iterative refinement in a structured manner. For the experiments in this section, we implement $\Phi$ using a RoBERTa-based model that was trained on the GoEmotions dataset~\cite{demszky2020goemotionsdatasetfinegrainedemotions}—a curated collection of Reddit comments that have been human-annotated for emotions. This model maps text to a 28-dimensional vector space representing unique emotions.

The process begins with a base open-source Large Language Model, $M_1$, which can be fine-tuned using a fine-tuning transcript, $F = (f_1, f_2, \dots, f_p)$.
Each interaction $f_i$ represents a user-model exchange, consisting of input-output pairs in the case of SFT, or response rankings in the case of RLHF.
A fine-tuning function $\mathcal{F}$ then updates the model based on this feedback, producing a refined version $M_{i+1}$. To quantify model performance, we define an ``ideal'' model $M_{ideal}$, which is assumed to generate ground-truth outputs for any input $x$. Then, the quality of a given model $M_i$ can be expressed as the distance between $\Phi(M_{ideal}(x))$ and $\Phi(M_i(x))$ in the embedding space. This distance represents how far the model's output is from the desired behavior. 

In our implementation, we define this emotion target by selecting the highest-ranking sentence for the given emotion from the GoEmotions dataset, retrieving its $k$-nearest neighbors, and averaging their vectors to create the target centroid. These $k$-nearest neighbors also serve as the fine-tuning transcript $F$. While our experiments use sentiment embeddings, this approach generalizes to any function providing meaningful vector representations of LLM outputs. Thus, we can quantify fine-tuning progress by measuring how model outputs move closer to the target in embedding space.

\subsection{Baseline: Single-Model Fine-Tuning} \label{sec:point_space/singlemodel}
To validate this conceptual framework, we apply our vector space approach to traditional single-model fine-tuning, where we feed a singular model a set of fine-tuning samples and examine whether it exhibits measurable convergence toward a defined target. These results serve as a baseline for Section~\ref{sec:point_space/multimodel}.

Starting with a base model, $M_1$, which is fine-tuned for a specific emotion using the $k$-nearest neighbors as the fine-tuning transcript, $F_i$, we assess how training data volume influences convergence by progressively increasing the number of samples in our fine-tuning set. Given a neutral prompt (``Write a Reddit comment''), the fine-tuned model generates responses that are mapped to the emotion space. The Euclidean distance between these response embeddings and the target centroid serves as our primary convergence metric.

\begin{figure}[htb]
    \centering
    \includegraphics[width=1\linewidth]{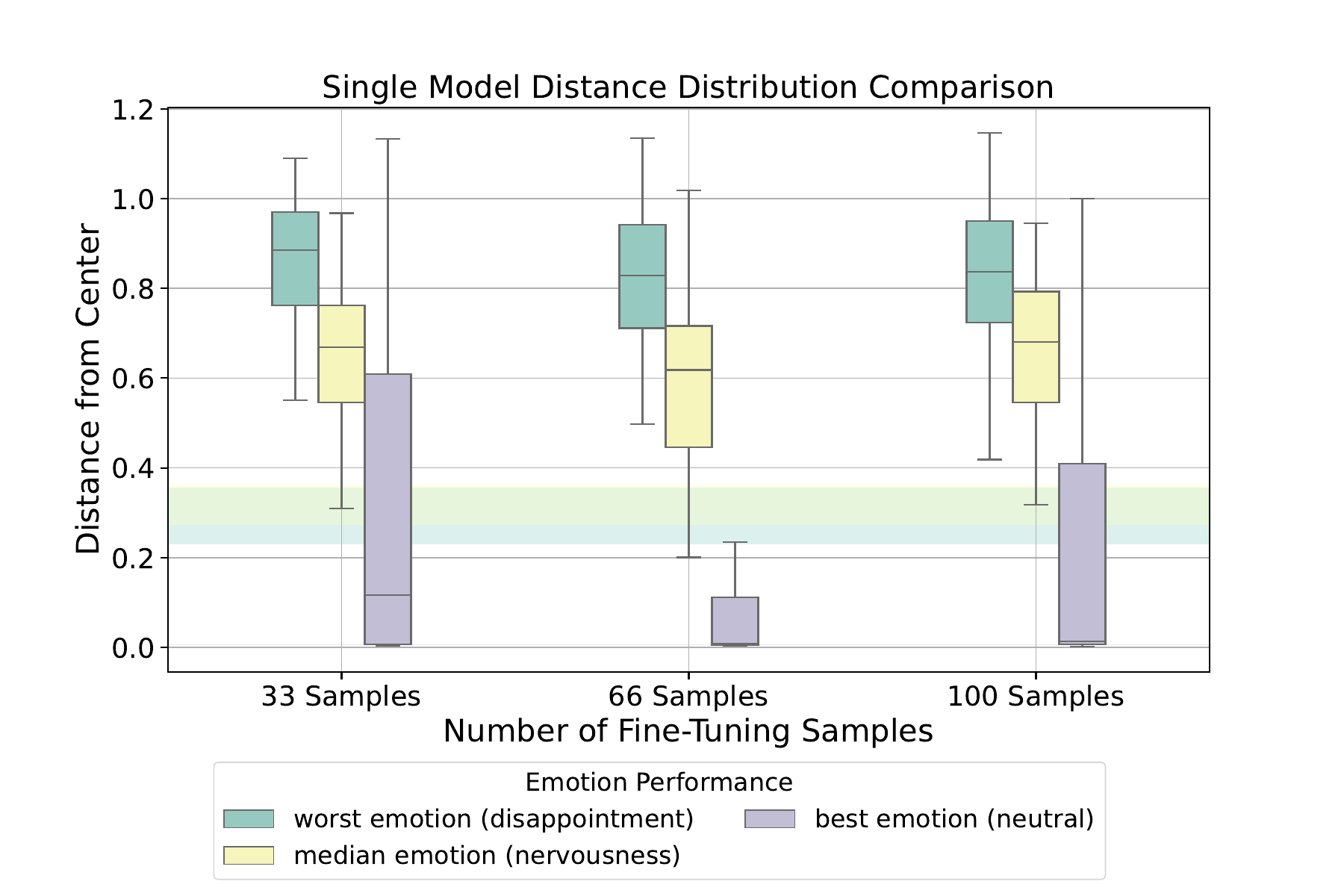}
    \caption{\textbf{Baseline single-model fine-tuning experiment.} This experiment demonstrates traditional fine-tuning and visualizes model convergence towards specific emotional vector spaces across varying sample sizes. Using the GoEmotions dataset of human-annotated Reddit comments, we identified the 100 nearest neighbors for each emotion's highest-ranking comment and fine-tuned independent models on 33, 66, and 100 samples per emotion. Box plots show results from prompting each model 50 times with ``Write a Reddit comment'' and processing outputs through the GoEmotions classifier. The y-axis represents distance to the target emotion centroid (mean of 100 nearest neighbors), with shaded bands showing the interquartile range of distances from these neighbor samples to the centroid. This graph displays the worst, median, and best-performing emotions based on mean distance to target centroid at 100 samples.}
    \label{fig:single-finetune}
\end{figure}

Figure~\ref{fig:single-finetune} illustrates the fine-tuning process across varying sample sizes for the best, median, and worst-performing models. These emotion models were ranked based on their outputs' mean Euclidean distance to the target centroid when trained on 100 samples. While there is some initial movement toward the target centroid, the improvement is modest, and after 66 samples, progress plateaus or even slightly deteriorates. This suggests that while expanding the fine-tuning set can nudge model outputs in the right direction, they may not be sufficient on their own to ensure consistent convergence. These limitations highlight the need for more robust fine-tuning strategies, which we propose in the following subsection.

\subsection{Foundational Framework: Multi-Model Selection Fine-Tuning}
\label{sec:point_space/multimodel}
Building on the single-model results, we introduce a competitive strategy that generates multiple model variants per iteration, selecting the best performer as the base for the next iteration. This competitive mechanism introduces selection pressure, ensuring that improvements accumulate more effectively over successive iterations.

Formally, at iteration $i$, we generate $j$ fine-tuned model variants, $M_i^{(j)}$, each trained on different subsets of feedback data $F_i^{(j)}$. For consistency with our single-model experiment, we use the same $k$-nearest neighbors for each emotion as the fine-tuning samples, setting $k=100$. The training samples are distributed evenly across all $j$ models in each iteration, with the winning model's training samples removed from the feedback pool in subsequent rounds. This progressive reduction in available training data ensures that each successive model learns from a unique set of samples. Each model produces a set of test responses using the generic prompt described in Section~\ref{sec:point_space/singlemodel}, which are then mapped to the same vector space as before. We then compute the distance between each model's response distribution and the target centroid, selecting the model that minimizes this distance as $M_{i+1}$.

\begin{figure}[htb]
    \centering
    \includegraphics[width=1\linewidth]{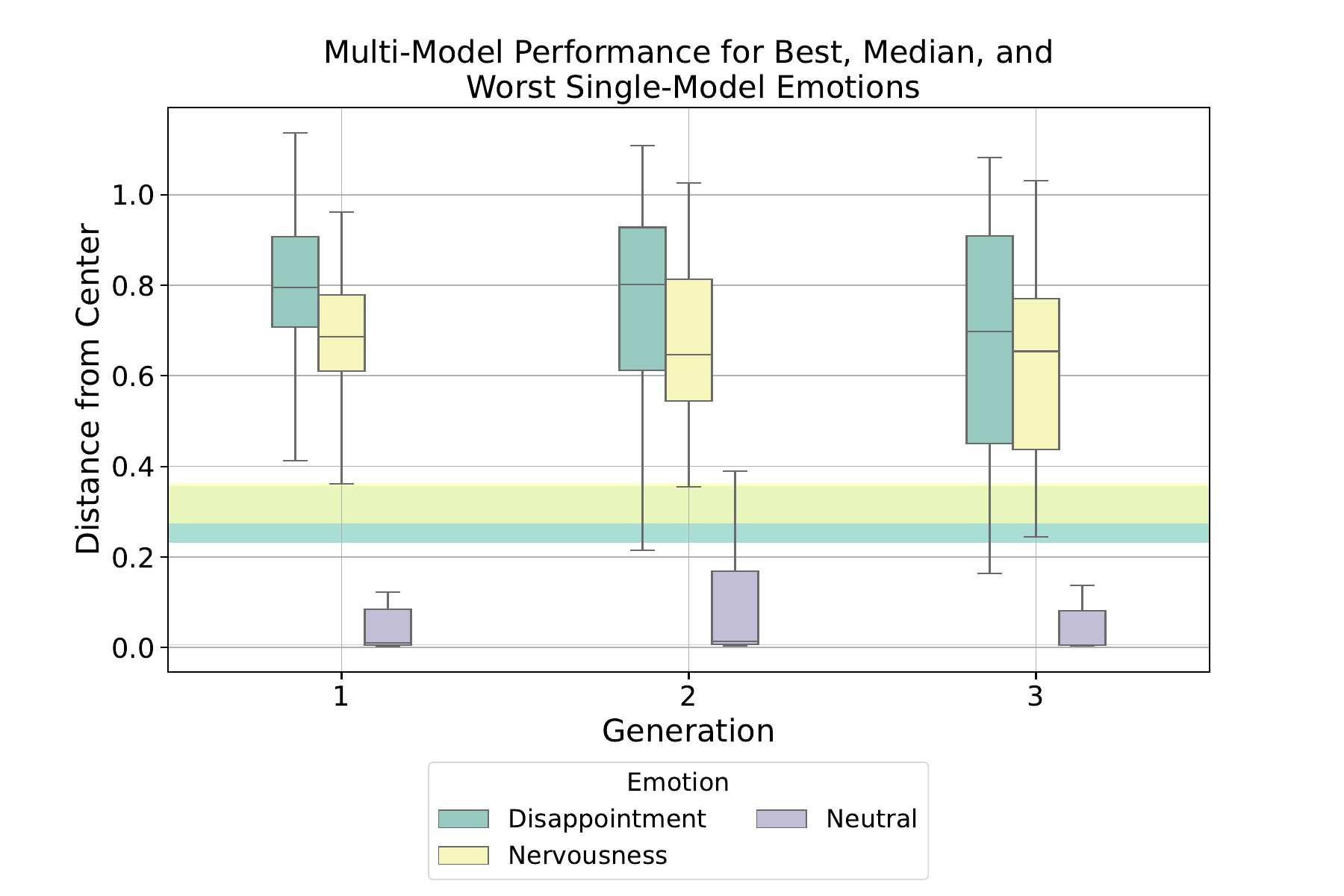}
    \caption{\textbf{Multi-model competitive fine-tuning performance.} This experiment implements a competitive fine-tuning approach using the same GoEmotions dataset, 100 nearest neighbor samples per emotion, and classifier methodology as Figure~\ref{fig:single-finetune}. The base model is cloned 3 times per iteration, with samples evenly split across clones for fine-tuning. After evaluation with the same generic prompt, the clone with smallest distance to the target emotion centroid becomes the winning model and base for the next iteration. The winning model's samples are removed from the training pool, ensuring subsequent models see unique data. Box plots show outputs from winning models for the same worst, median, and best-performing emotions from Figure~\ref{fig:single-finetune}. Compared to the single-model approach using all 100 samples, the competitive approach in the third iteration reduced mean distances by 20.04\%, 5.24\%, and 55.37\%, respectively.}
    \label{fig:multi-finetune}
\end{figure}

Figure~\ref{fig:multi-finetune} presents the results of our multi-model selection approach, evaluated using the best, median, and worst-performing emotions identified in Figure~\ref{fig:single-finetune}. Compared to the single-model approach fine-tuned on 100 samples, our multi-model selection method, after three iterations, reduced the mean distance to the target centroid by 20.04\% for \emph{disappointment}, 5.24\% for \emph{nervousness}, and 55.37\% for \emph{neutral}. These results indicate that fine-tuning under a selection-driven process not only improves contribution tracking but also enhances overall model quality through competitive refinement.

\begin{figure}[htb]
    \centering
    \includegraphics[width=1\linewidth]{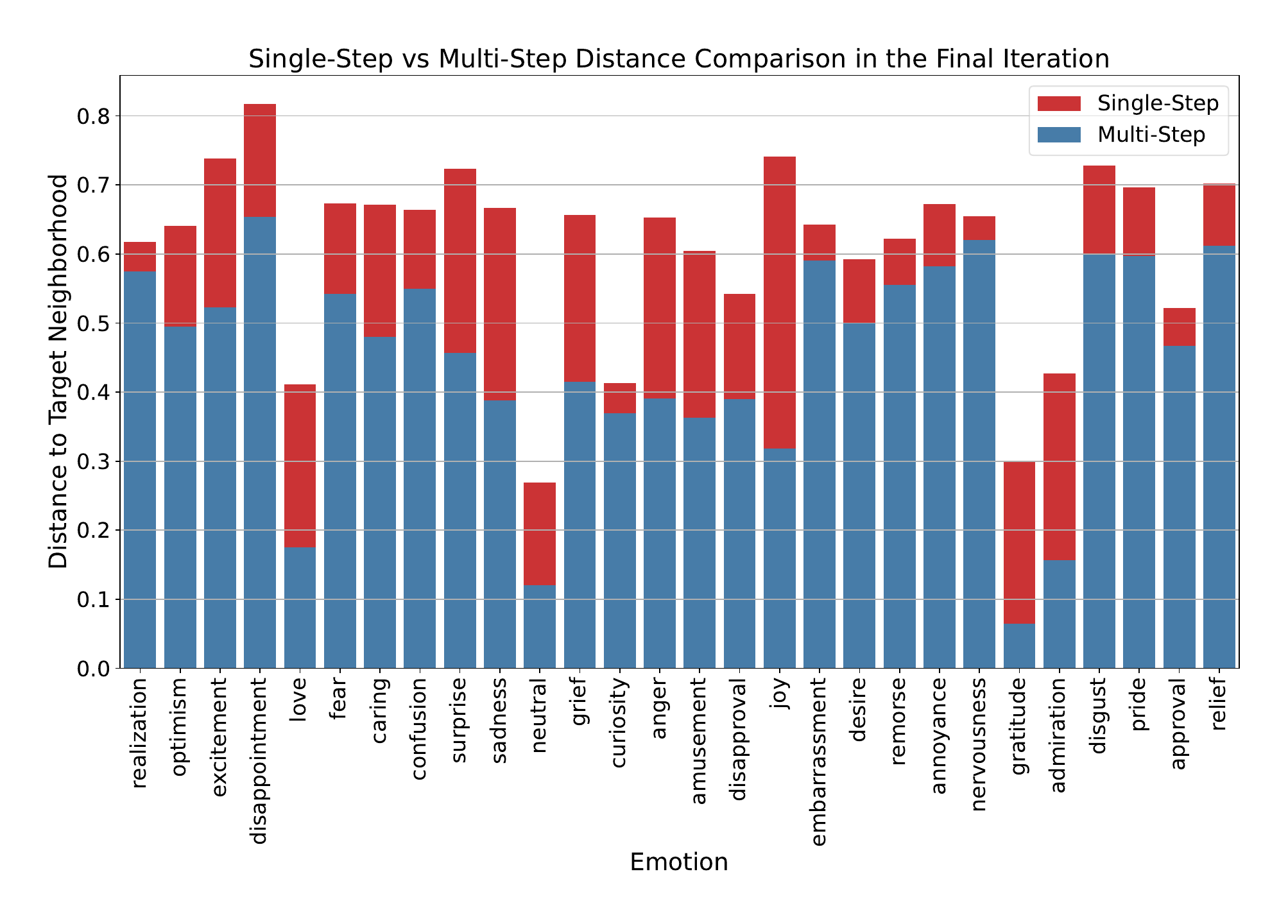}
    \caption{This figure compares the performance of single-step fine-tuning at 100 samples from Figure~\ref{fig:single-finetune} with multi-step competitive fine-tuning in the third iteration from Figure~\ref{fig:multi-finetune} across all 28 emotions in the GoEmotions dataset. Each stacked bar shows the distance to the target emotion centroid, with red representing single-step performance and blue representing multi-step performance. The multi-step approach demonstrates superior performance across all emotions, with particularly notable improvements in emotions like \emph{love}, \emph{gratitude}, and \emph{admiration}.}
    \label{fig:comparison}
\end{figure}

To provide a comprehensive view of the performance differences, Figure~\ref{fig:comparison} extends this comparison across all target emotions. The results confirm that multi-model fine-tuning outperforms single-model fine-tuning consistently, showcasing that our competitive framework not only improves model performance but also enables the contribution tracking mechanisms outlined in the following section.

\section{Experiment 2: Collaborative Fine-Tuning Framework} \label{sec:grouping_eval_contr}
Having established the viability of our multi-model, competitive fine-tuning approach, we now investigate different strategies for grouping users and evaluating model performance. Specifically, we aim to determine how different grouping methods impact convergence and whether our contribution-tracking mechanism reliably identifies user influence. To do so, we experiment with multiple ways of partitioning users into training groups and compare several distance-based evaluation metrics for selecting the best-performing model at each iteration.

We introduce a point-based system to estimate individual user contributions. To assess the fairness of this mechanism, we compare each user's accumulated points to their approximate Shapley value. Since calculating exact Shapley values is computationally expensive, we leverage the KernelSHAP algorithm~\cite{10.5555/3295222.3295230,pmlr-v130-covert21a}, implemented in the \texttt{SHAP} Python library, to efficiently approximate these values.

\subsection{Definitions: Grouping and Evaluation Methods}
At each iteration, users are divided into groups, each fine-tuning a separate model. The model that best aligns with an expert-defined target becomes the base model for the next iteration, and all users receive rewards based on their group's performance. We test three grouping strategies and three evaluation metrics:

\begin{itemize}
    \item \textbf{Grouping Methods ($G_{method}$):}
    \begin{enumerate}
        \item \emph{Random Grouping:} Users are randomly assigned to groups with equal probability.
        \item \emph{$\epsilon$-greedy:} Initially, users are randomly grouped, but over time, assignment increasingly favors users with higher accumulated points, inspired by the popular $\epsilon-$greedy multi-armed bandit algorithm.
        \item \emph{Interleaved Grouping:} Users are first ranked based on prior contributions, then divided into high and low-performing groups. Assignments alternate between these two groups to create balanced teams.
    \end{enumerate}
    \item \textbf{Evaluation Methods ($E_{method}$):}
    \begin{enumerate}
        \item \emph{L2 Norm (Euclidean Distance)}
        \item \emph{L1 Norm (Manhattan Distance)}
        \item \emph{Dot Product}
    \end{enumerate}
\end{itemize}

\begin{figure}[h]
    \centering
    \begin{tikzpicture}[
        node distance=1.0cm and 2.0cm,
        every node/.style={text centered, align=center, font=\footnotesize},
        user/.style={circle, draw=blue, fill=blue!10, minimum size=0.8cm},
        model/.style={rectangle, draw=black, fill=gray!10, minimum width=2cm, minimum height=1cm},
        process/.style={rectangle, draw=red, fill=red!10, minimum width=2.5cm, minimum height=1cm},
        arrow/.style={thick, ->, >=stealth}
    ]

    \node[user] (U1) {User 1};
    \node[user, right=0.2cm of U1] (U2) {User 2};
    \node[text width=0.5cm, right of=U2] (U3) {\Large ...};
    \node[user, right of=U3] (U4) {User $n$};

    \node[process, below=1.5cm of U2.east, xshift=0.5cm] (Grouping) {$G_{method}$};

    \node[model, below=0.8cm of Grouping, xshift=-1.8cm] (M1) {Model $M_1$};
    \node[right=0.3cm of M1] (Mdots) {\Large ...};
    \node[model, below=0.8cm of Grouping, xshift=1.5cm] (M2) {Model $M_m$};

    \node[process, below=2.5cm of Grouping] (Selection) {Select Best Model based on $E_{method}$};

    \node[process, below=0.5cm of Selection] (Update) {Update User Scores};

    \draw[arrow] (U1.south) -- (Grouping.north);
    \draw[arrow] (U2.south) -- (Grouping.north);
    \draw[arrow] (U4.south) -- (Grouping.north);

    \draw[arrow] (Grouping.south) -- (M1.north);
    \draw[arrow] (Grouping.south) -- (M2.north);

    \draw[arrow] (M1.south) -- (Selection.north);
    \draw[arrow] (M2.south) -- (Selection.north);

    \draw[arrow] (Selection.south) -- (Update.north);

    \draw[arrow, dashed] 
        (Update.south) -- ++(0,-0.5) 
        -- ++(-3.5,0)  
        -- ++(0,6.05)  
        -- (Grouping.west); 

    \end{tikzpicture}
    \caption{\textbf{Iterative Decentralized Fine-Tuning Process}: This figure illustrates our framework where multiple users contribute to model fine-tuning across iterative rounds. Users are grouped using $G_{method}$ and assigned to fine-tune separate models ($M_1$ to $M_m$). After each iteration, models are evaluated using $E_{method}$ to select the best-performing model. User scores are updated based on their group's model performance, and the winning model serves as the base for the next iteration, creating a continuous cycle.}
    \label{fig:fine_tuning_process}
\end{figure}
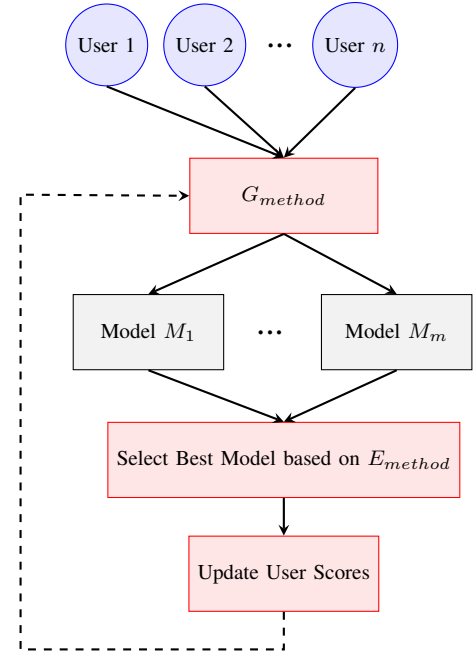

\subsection{Simulation Setup}\label{sec:simulation-setup}
We use the MovieLens dataset\footnote{\url{https://grouplens.org/datasets/movielens/}}, which contains user preference data across $n$ different movie genres, treating each genre as an independent dimension in our vector space. Figure~\ref{fig:fine_tuning_process} shows a high-level overview of the simulation, and is structured as follows:

\begin{enumerate}
    \item \textbf{Initialization:} A set of users and an initial model $M_1 \in \mathbb{R}^n$ are randomly placed in the feature space. We define an expert target ($\mathcal{E}$) as the preference vector of a single randomly selected user.
    \item \textbf{Grouping:} Users are divided into groups based on $G_{method}$, and each group fine-tunes a model instance.
    \item \textbf{Model Update:} For each group, we compute a weighted centroid $\mathbf{C}_{g}(t)$, where each user's preference vector is weighted by their accumulated point value at round $t$. The candidate model for group $g$ is updated as:
    \[
    \mathbf{M}_{\text{candidate}, g}(t) = \mathbf{M}_t + \delta \left( \mathbf{C}_{g}(t) - \mathbf{M}_t \right)
    \]
    where $\mathbf{M}_t$ is the base model for round $t$.
    \item \textbf{Selection:} The candidate model closest to the expert target—determined by $E_{method}$—becomes base model in round $t+1$. To introduce noise, we arbitrarily assume the expert selects an incorrect model 5\% of the time.
    \item \textbf{Rewarding Contributions:}  
    At each round, we rank the candidate models based on their proximity to the expert target, using \(E_{method}\). Suppose there are $m$ groups. The best-performing group's model receives \(m\) points, the second-best receives \(m-1\), and so on, down to \(1\) point for the lowest-ranked model. Formally, each user in group $g$ receives $\Delta_g$ points, where
    $$
    \Delta_g = m - \text{argsort} \left( E_{method}(M_{\text{candidate}, g}(t), \mathcal{E}) \right) .
    $$
    \item \textbf{Final Evaluation:} At the end of 100 rounds, we compute the estimated Shapley values for each user and compare them to their accumulated points to measure correlation.
\end{enumerate}

\begin{figure}[htb]
    \centering
    \includegraphics[width=1\linewidth]{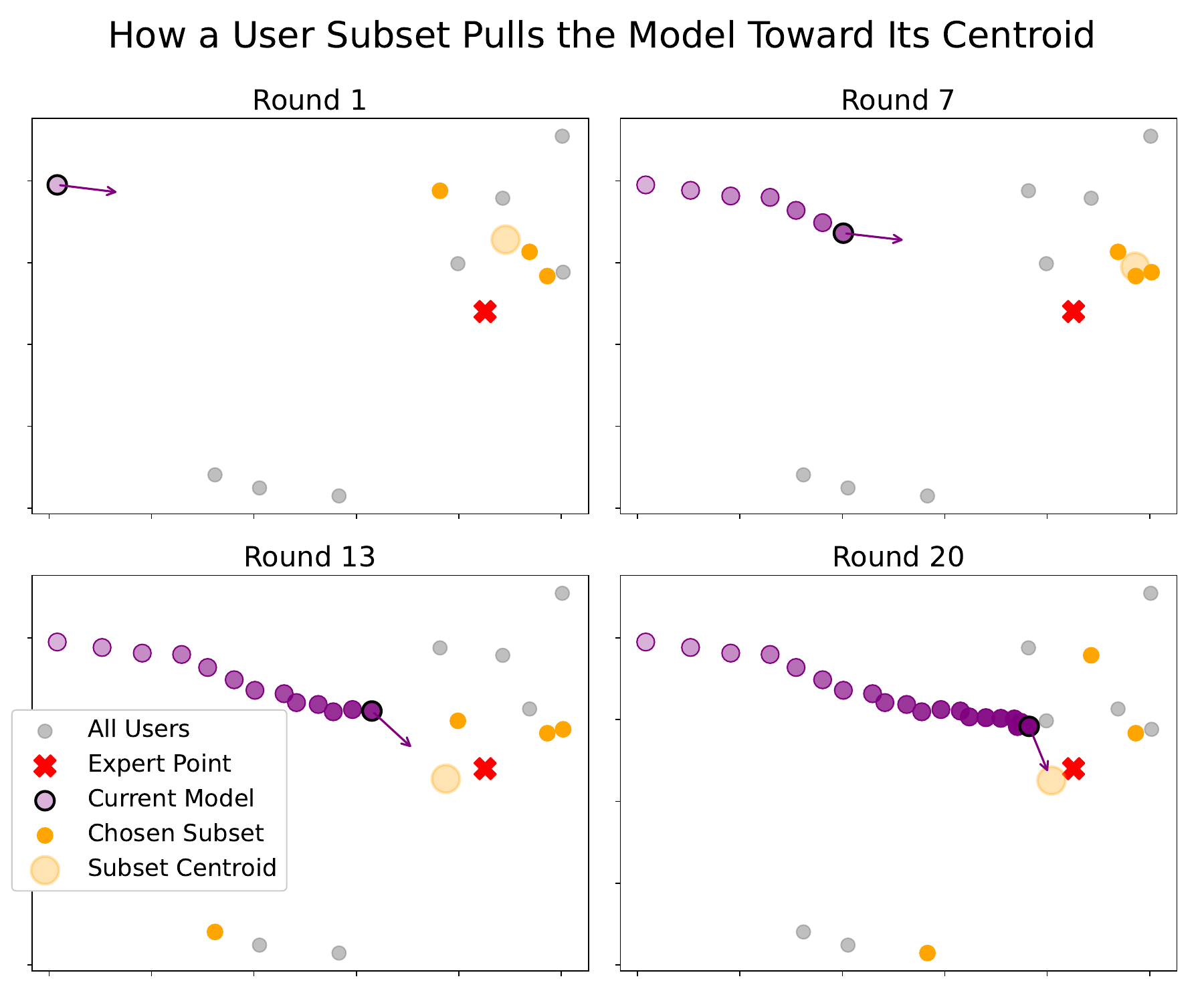}
    \caption{\textbf{User Subset Impact on Model Convergence}: This figure demonstrates the iterative convergence process, where user preferences are represented as points in a multi-dimensional space. In each round, a selected user subset (dark yellow points) computes its weighted centroid (big yellow point), which pulls the current model (purple circle with black outline) toward the expert target (red X). The arrows show the directional influence of each subset's centroid on model updates. Over 20 rounds, the model gradually converges toward the expert point as different user subsets contribute to fine-tuning, illustrating how decentralized user contributions can guide model alignment toward desired outcomes.}
    \label{fig:subset_influence}
\end{figure}

The intuition behind this approach is that each group's model update shifts the model toward the weighted centroid of its users' preferences, mimicking how subgroups of annotators might fine-tune a model based on their distinct biases and perspectives. Figure~\ref{fig:subset_influence} provides a conceptual illustration of this process in action. In the figure, the arrow shows the directional influence of the computed user subset centroid on the current model point. Over successive rounds, where only the model closest to the expert-defined target is retained, the model gradually converges toward the expert point. The deliberate introduction of noise in the selection process accounts for inconsistencies that might arise in real-world human evaluations, ensuring that our method is robust to occasional misjudgments. Finally, tracking user contributions through accumulated rewards allows us to assess whether this competitive structure fairly credits influence to the most impactful participants.

\subsection{Results: Performance of Grouping-Evaluation Method Combinations}
We run the simulation 50 times to compute the average performance of each $(G_{method}, E_{method})$ combination, and track:
\begin{itemize}
    \item \textbf{Convergence Distance:} The final Euclidean distance to the expert target.
    \item \textbf{Pearson Correlation:} The relationship between user point accumulation and their estimated Shapley values.
\end{itemize}

\begin{figure}[htb]
    \centering
    \includegraphics[width=1\linewidth]{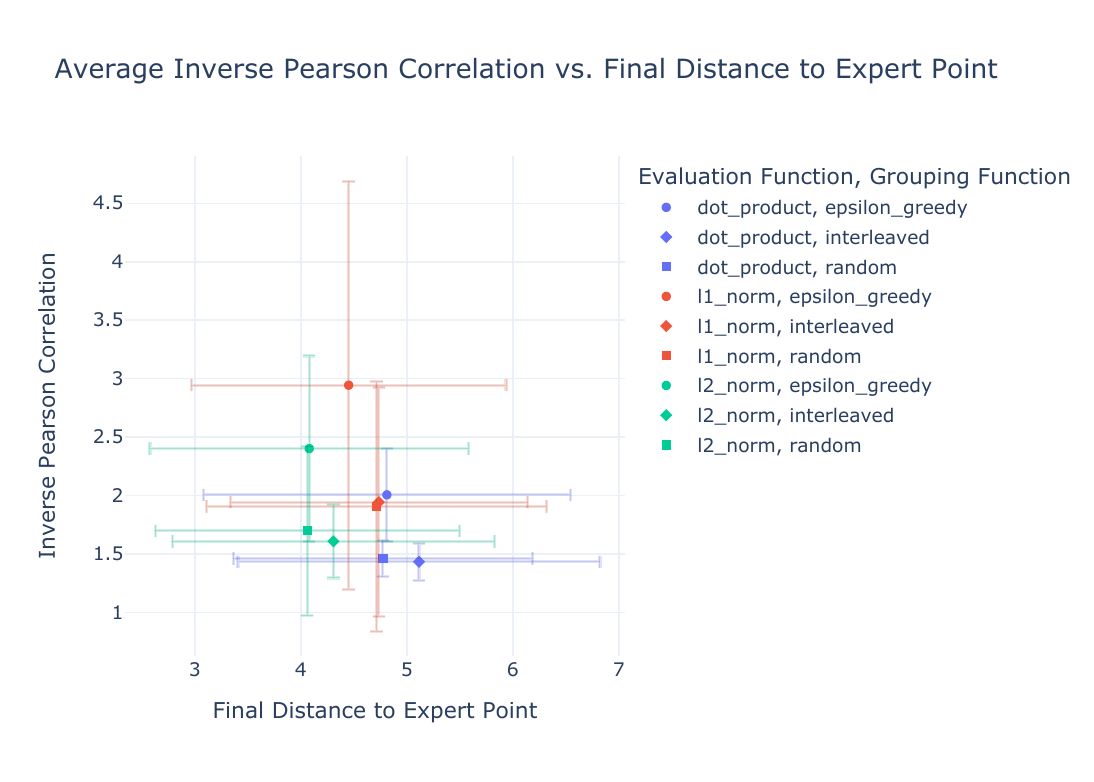}
    \caption{\textbf{Evaluation of Method Combinations in Collaborative Model Training}: This figure shows simulation results for the framework outlined in Section~\ref{sec:simulation-setup}. Using the MovieLens dataset of user preferences toward movie genres to represent humans in the loop, we randomly sample users by their preference vectors and select a random expert target. The experiment tests combinations of grouping and evaluation methods, rewarding users and updating the model toward each round's winning subset centroid. Results show average performance across 50 simulation runs with 50 users and 3 groups. Each point represents the trade-off between model convergence to the expert target (x-axis) and reward system accuracy compared to Shapley values (y-axis). Lower values indicate better performance; error bars show standard deviation.}
    \label{fig:pearsondist}
\end{figure}

We compute the Pearson correlation between accumulated user scores in the final round and corresponding Shapley values to assess our point-based method's accuracy. The Pearson correlation coefficient measures the strength of the linear relationship between two variables. A coefficient near 1 indicates our system closely approximates theoretical Shapley values, rewarding users proportionally to their actual impact. Lower correlation indicates deviation from ideal tracking, suggesting users are improperly rewarded relative to their actual influence.

We found that deterministic grouping approaches produce lower Pearson correlations, suggesting unfair reward distribution relative to true user contributions. This likely occurs because deterministic strategies limit the redistribution of users over time, reinforcing early biases in contribution estimates. In contrast, methods incorporating randomness lead to higher Pearson correlations, meaning that user impact is better captured and rewarded more fairly. Additionally, these methods also improve model convergence, indicating structured randomness aids both model optimization and accurate identification of impactful contributors.

Figure~\ref{fig:pearsondist} showcases an initial analysis comparing unique pairs of $(G_{method}, E_{method})$ based on the inverse Pearson coefficient and final convergence distance for 50 users and 3 groups. Since lower values are better for both metrics, the best-performing method combinations appear closer to the origin. In this configuration, L2 Norm + Random delivers the best convergence distance, while Dot Product + Interleaved achieves the best inverse Pearson coefficient. These results show that L2 Norm performs best for convergence distance while Dot Product does best for inverse Pearson coefficient.

\subsection{Results: Effects of User and Group Scaling on Method Performance}
\begin{figure}[htb]
    \centering
    \includegraphics[width=1\linewidth]{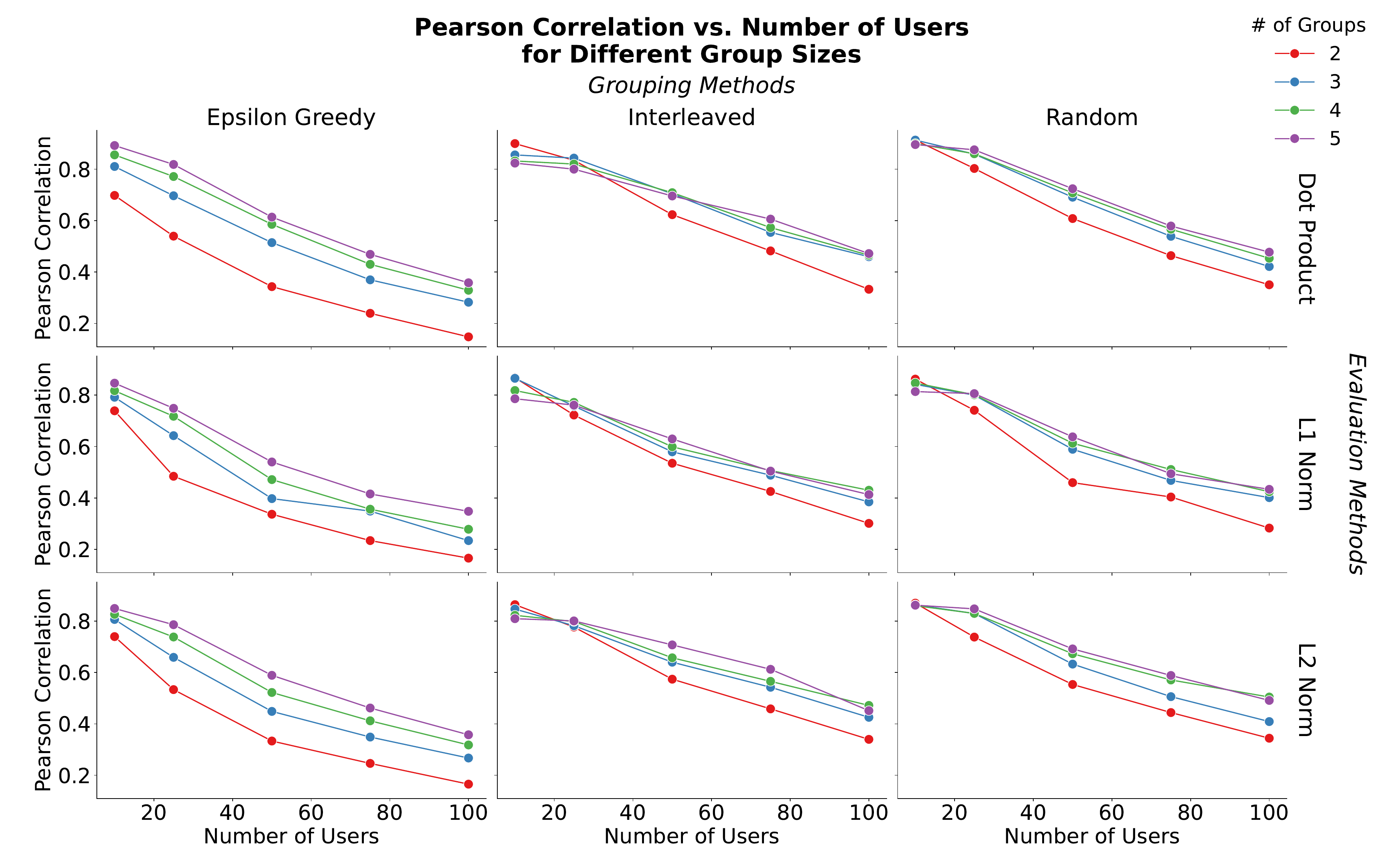}
    \caption{\textbf{Scaling Effects on Reward Fairness}: This figure shows a holistic view of the scaling factors in the experiment outlined in Section~\ref{sec:simulation-setup}. We examine how increasing the number of users from 10 to 100 (x-axis) and varying group sizes from 2 to 5 (colored lines) affect the accuracy of our reward system in estimating user contributions. The y-axis represents Pearson correlation between accumulated user points and actual Shapley values in the final round, averaged across 50 simulation runs. Results are shown for all nine combinations of grouping methods (columns) and evaluation methods (rows). The consistent downward trend across all methods indicates that reward accuracy decreases as user populations grow, while higher group counts generally maintain better correlation.}
    \label{fig:corr_users_groups}
\end{figure}

\begin{figure}[htb]
    \centering
    \includegraphics[width=1\linewidth]{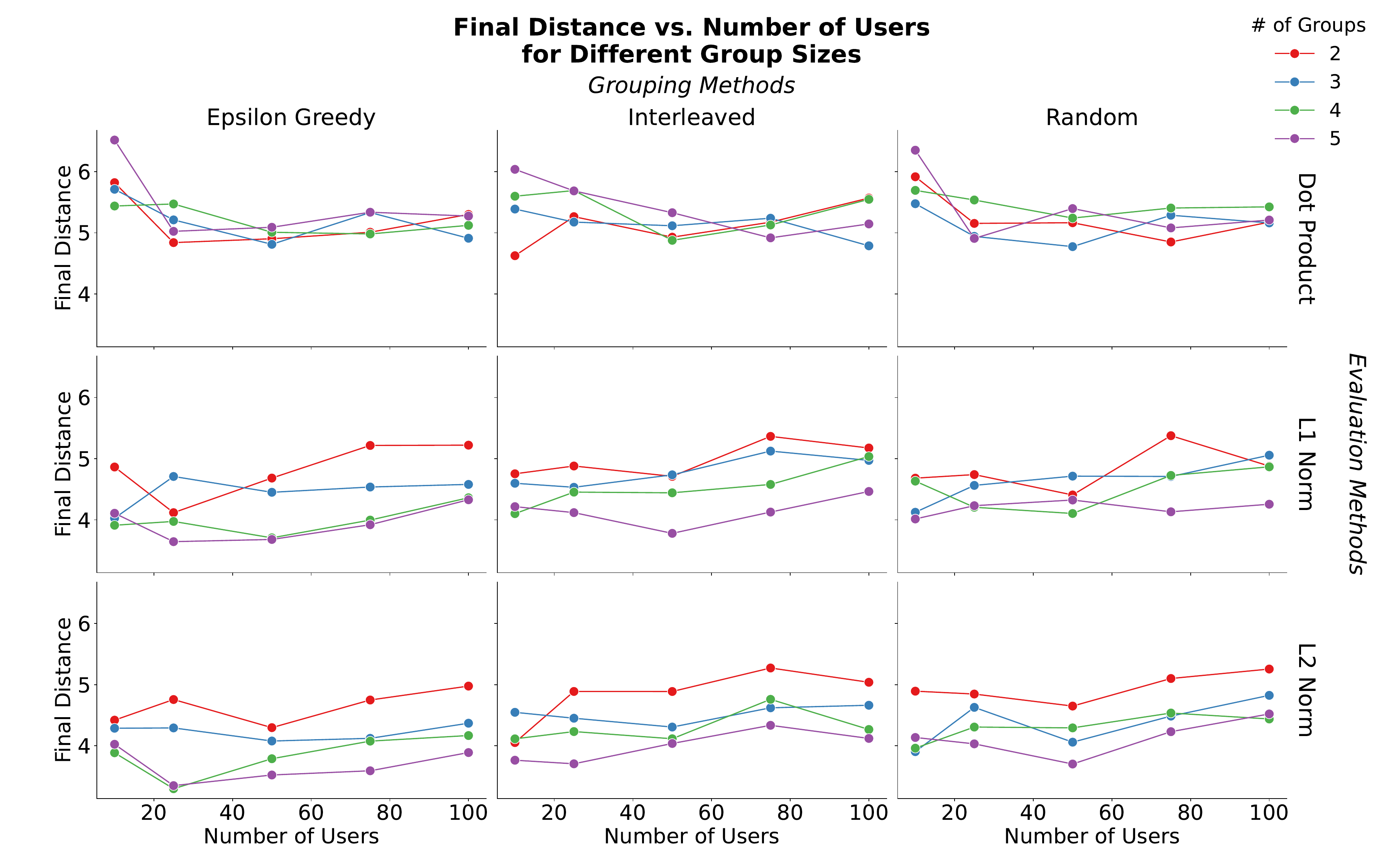}
    \caption{\textbf{Scaling Effects on Model Convergence}: This figure examines the same scaling factors as Figure~\ref{fig:corr_users_groups}, but focuses on how scaling impacts model convergence to the expert target. The y-axis represents final distance to the expert point, averaged across 50 simulation runs, with results shown for all nine combinations of grouping methods (columns) and evaluation methods (rows). Unlike the consistent decline in reward accuracy shown in Figure~\ref{fig:corr_users_groups}, convergence distance remains relatively stable across user and group scaling.}
    \label{fig:dist_users_groups}
\end{figure}

The ability of an open, collaborative system to scale effectively is crucial for robustness. To investigate this, we systematically explore the parameter space by varying the number of users ($\in\{10, 25, 50, 75, 100\}$) and groups ($\in [2-5]$), examining how each $(G_{method}, E_{method})$ pair responds to these scaling challenges. Figures~\ref{fig:corr_users_groups} and~\ref{fig:dist_users_groups} show the average performance across the 50 simulation runs for Pearson Correlation and convergence distance, respectively, across the entire parameter space. 

Figure~\ref{fig:corr_users_groups} shows Pearson correlation declining as user numbers increase, regardless of grouping or evaluation method. While our naive reward mechanism struggles to approximate Shapley values as user populations grow, meaningful correlation persists, especially in smaller settings where user contributions are easier to estimate. Higher group counts yield better Pearson coefficients, yet this improvement remains secondary to the correlation decline caused by increasing user counts.

In contrast to the more pronounced decline in Pearson correlation, the final distance remains relatively stable as the number of users grows, as seen in Figure~\ref{fig:dist_users_groups}. The real impact in convergence distance stems from differences in evaluation methods—L2 Norm generally achieves lower distances. Dot Product shows poor convergence with fewer users but improves as user count increases, while L1 and L2 Norm achieve better convergence with smaller user populations before gradually deteriorating with higher user counts.

\begin{table}[ht]
    \centering
    \caption{Top 3 method pairs achieving the best convergence distance and Pearson correlation, with winning percentages aggregated across all parameter configurations.}
    \begin{tabular}{|p{1.75cm}|p{1.75cm}|r|}
    \hline
    \multicolumn{3}{|c|}{\textbf{Best Convergence Distance}} \\
    \hline
    \textbf{Grouping Function} & \textbf{Evaluation Function} & \textbf{Winning Percentage} \\
    \hline
    $\epsilon$-Greedy & L2 Norm  & 60\% \\
    \hline
    $\epsilon$-Greedy & L1 Norm  & 15\% \\
    \hline
    Random & L2 Norm  & 10\%  \\
    \hline
    \hline
    \multicolumn{3}{|c|}{\textbf{Best Pearson Correlation}} \\
    \hline
    \textbf{Grouping Function} & \textbf{Evaluation Function} & \textbf{Winning Percentage} \\
    \hline
    Random & Dot Product & 45\% \\
    \hline
    Interleaved & Dot Product & 40\% \\
    \hline
    Random & L2 Norm  & 10\%  \\
    \hline
    \end{tabular}
    \label{tab:combined_stacked}
\end{table}

Table~\ref{tab:combined_stacked} compares which method pairs consistently outperform others across the parameter space. It reveals that Dot Product evaluation (with Random or Interleaved) dominates in Pearson correlation, achieving the best performance in 85\% of all user-group configurations. In contrast, we can see that the $\epsilon-$Greedy + L2 Norm pair minimizes convergence distance, winning in 60\% of configurations. These findings reveal a trade-off between reward accuracy and model convergence, with optimal methods depending on system priorities.

\section{Conclusion} \label{sec:conclusion}
We introduced a novel framework for decentralized fine-tuning of LLMs that leverages a structured, competitive SFT process. Our approach enables a broader pool of users to iteratively refine a model while tracking individual contributions. By dividing users into groups, fine-tuning multiple competing models, and selecting the best-performing variant in each iteration, we optimize both model quality and fair contribution assessment.

These empirically validated fine-tuning experiments, framed as movement through an $n$-dimensional vector space, demonstrated that model outputs systematically improve when guided by a structured selection mechanism, leading to reliable model convergence. We also explored various grouping strategies and evaluation metrics, analyzing their effects on convergence accuracy and the fairness of contribution tracking. Our findings suggest that incorporating randomness into grouping methods improves both model performance and the reliability of user contribution, with L2 Norm evaluation yielding the best convergence distance and Dot Product providing the most reliable estimation of user contributions through Pearson correlation.

This framework revealed that competition serves as a contribution tracking mechanism and as an intrinsic driver of improved performance. Iteratively selecting best-performing models creates a selection pressure that accelerates alignment with desired outcomes. Overall, our results demonstrate that a decentralized, competitive SFT approach is feasible and beneficial, enabling more scalable and inclusive LLM fine-tuning methods. Leveraging diverse collective intelligence creates models that better reflect human preferences while ensuring fair rewards for meaningful contributions.

\section{Future Work} \label{sec:future}
While this work provides a strong proof of concept, several directions remain for future exploration. A critical next step is validating the framework in real-world settings with human participants, such as crowd-sourced annotators or domain experts. This would test the system’s practical feasibility in more diverse and variable environments. Our current approach assumes honest and consistent participation, but future work should explore how inconsistent or adversarial behavior affects model performance and contribution tracking fairness. Future research could also examine the impact of varying levels of evaluator error (currently set at 5\% in Section~\ref{sec:simulation-setup}) on system performance. Modeling such behaviors and evaluating system resilience under these conditions will be crucial for deployment in realistic environments.

Additionally, the reward system could be extended to support more granular or alternative mechanisms. As the number of users scales, the current winner-takes-most approach may lead to reduced fairness or diminished incentive to contribute. Exploring modifications to the reward structure may yield fairer and more motivating outcomes, particularly in larger groups.

\section*{Acknowledgment}
Sections of this report have been copy-edited with the assistance of ChatGPT. We certify that ChatGPT was not utilized to produce any technical content and we accept full responsibility for the content of the paper.

\bibliographystyle{IEEEtran}
\bibliography{references}

\end{document}